\documentclass[%
 reprint,
 amsmath,amssymb,
 aps,
prb,
]{revtex4-1}

\usepackage{graphicx}% Include figure files
\usepackage{dcolumn}% Align table columns on decimal point
\usepackage{bm}% bold math
\usepackage{dcolumn}
\usepackage{color, colortbl}
\definecolor{c1}{RGB}{185, 184, 217}
\definecolor{c2}{RGB}{217, 184, 203}
\definecolor{c3}{RGB}{217, 212, 184}
\definecolor{c4}{RGB}{184, 217, 193}
\newcolumntype{s}{>{\columncolor{c1}}d}
\newcolumntype{w}{>{\columncolor{c1}}c}
\newcolumntype{t}{>{\columncolor{c2}}d}
\newcolumntype{x}{>{\columncolor{c2}}c}
\newcolumntype{u}{>{\columncolor{c3}}d}
\newcolumntype{y}{>{\columncolor{c3}}c}
\newcolumntype{v}{>{\columncolor{c4}}d}
\newcolumntype{z}{>{\columncolor{c4}}c}

\begin{document}

\preprint{APS/123-QED}

\title{Quantitative metamaterial property extraction}% Force line breaks with \\

\author{David Schurig}
\author{Alex Orange}

\affiliation{Department of Electrical and Computer Engineering, University of Utah, Salt Lake City, UT, 84112, USA}

\date{\today}% It is always \today, today,

\begin{abstract}
We examine an extraction model for metamaterials, not previously reported, that gives precise, quantitative and causal representation of S-parameter data over a broad frequency range, up to frequencies where the free space wavelength is only a modest factor larger than the unit cell dimension. The model is comprised of superposed, slab-shaped response regions of finite thickness, one for each observed resonance. The resonance dispersion is Lorentzian and thus strictly causal. This new model is compared with previous models for correctness likelihood, including an appropriate Occam’s factor for each fit parameter. We find that this new model is by far the most likely to be correct in a Bayesian analysis of model fits to S-parameter simulation data for several classic metamaterial unit cells.
\end{abstract}

\maketitle

\section{INTRODUCTION}
The field of metamaterials has promised a dramatically expanded range of material properties to the electromagnetic designer. However, the compelling performance gains that could be realized in many devices, are tempered by two problems. One is that the desirable range and tunability of the real parts of the permittivity and permeability come also with undesirable qualities: loss and spatial dispersion. Even quantifying these undesirable qualities is a challenge. Not only is there no simple standard metric for quantifying spatial dispersion, but it’s presence makes dubious the practice of quantifying loss with material property imaginary parts.

The second problem is that we lack robust algorithms for assessing a given metamaterial design without ad-hoc human intervention. Ideally, our algorithm would provide simple and physically meaningful, quantitative descriptions of the effective behavior. Without such algorithms one cannot hope to exploit the large scale computational resources and optimization strategies that would lead to superior designs.

The premise of the current work is that extracting metamaterial properties from simulation data into the best model will provide progress with both of these problems. For the purpose of this article, the best model is the simplest one that provides accurate quantitative representation of the simulated behavior. Significant spatial dispersion seems to be unavoidable with metamaterials that provide their unique (and sometimes extreme) properties with practical unit cell dimensions. Several authors have suggested that a spatially dispersive model provides more physical insight than a spatially local one\cite{smith2010Analytic,Alu2011First,Alu2011Restoring,Alu2011Causality,Liu2013Generalized}. In this work we describe several models that incorporate spatial dispersion through unit cell inhomogeneity. We analyze these inhomogeneous models - including some not previously discussed for metamaterials - in an objective statistical analysis to identify a preferred model for several typical unit cell designs.

All the extraction models presented here are comprised of homogeneous, isotropic, layers, for which the reflection and transmission coefficients of normally incident plane waves can be analytically calculated. A metamaterial unit cell is simulated and its normal incidence reflection and transmission coefficients computed. Model parameters are found by fitting reflection and transmission coefficients from the model to those of the simulation data. These model parameters include the thicknesses of superposed, slab-shaped response regions, and their corresponding Lorentzian susceptibilities. The fits are performed over the entire frequency range simultaneously\cite{Lubkowski2007Extraction}.

Of note, we do not incorporate spatial dispersion or magneto-electric coupling into the slabs response itself, though in a homogenized sense, the layered structures can represent behavior that appears as such. If the models perform well, then adding such complexity is unnecessary\cite{smith2010Analytic}. 

The simulations presented here have been performed with the FDTD solver of CST Microwave studio, but other solvers should provide similar results.

\section{THE FOUR MODELS}
\begin{figure}
\includegraphics[width = 74 mm]{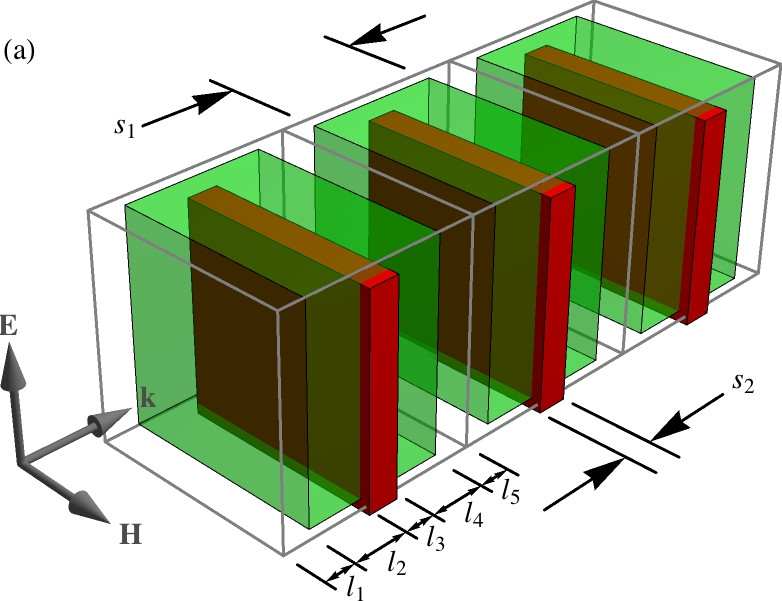}
\includegraphics[width = 74 mm]{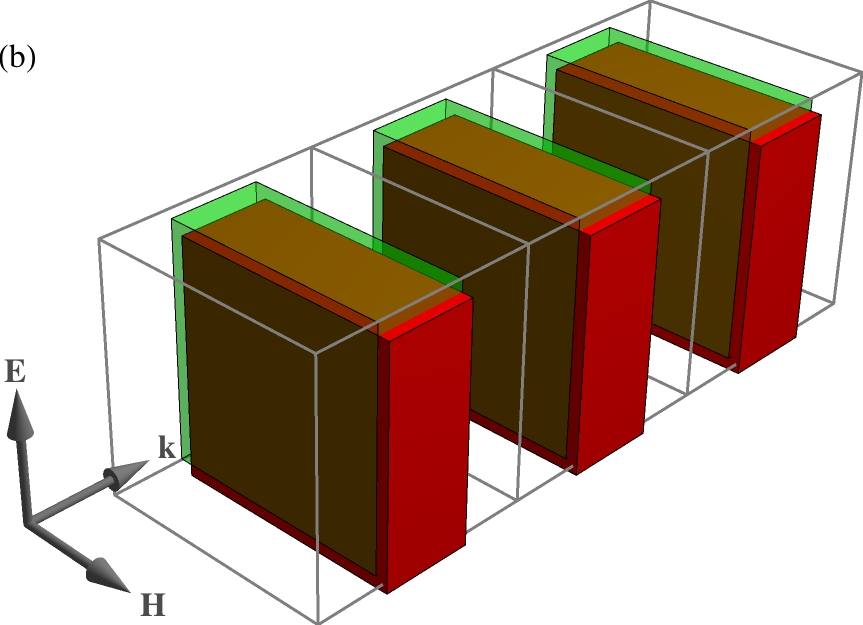}
\includegraphics[width = 74 mm]{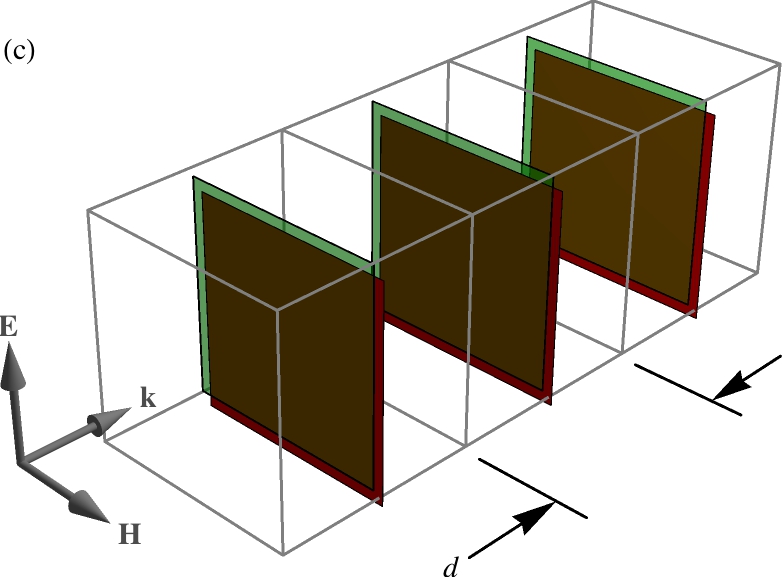}
\includegraphics[width = 74 mm]{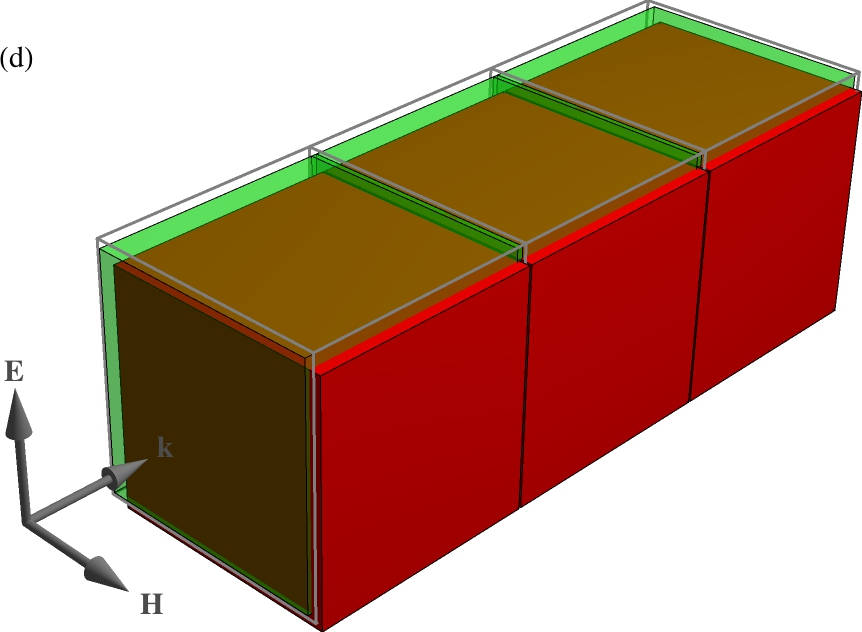}
\caption{\label{modelsFig}The four models: (a) multi-thickness, (b) single-thickness, (c) thin sheet, and (d) homogeneous. Two slab-shaped response regions (red and green) are shown, but an arbitrary number are allowed, one for each resonance. (Slabs are of infinite transverse extent.)}
\end{figure}
In describing the extraction models we will analyze, we will define the concepts of \textit{response slabs} and \textit{material layers} in a specific way. Geometrically, both constructs are assumed to be infinite in extent in the directions transverse to the wave propagation direction. \textit{Response slabs} have a thickness and a spatially uniform susceptibility (either electric or magnetic) arising from a single resonant mode. Since the total response at any given point in space is the sum of the responses due to all resonant modes that extend to that point, we represent the total response by a superposition of response slabs. This superposition of slabs creates a number of distinct uniform \textit{material layers} with uniform material properties. In a given layer, these properties are just the sum of the susceptibilities of the slabs that overlap that layer. For example, in FIG.~\ref{modelsFig}(a), the unit cell has two overlapping slabs, but five material layers. Two outer layers have vacuum properties (zero electric and magnetic susceptibility). Moving inward, there are two layers with the material properties give by only the susceptibility of the green slab, and a single, central layer with material properties equal to the sum of the susceptibilities of the green \textit{and} red slab.
For this work we assume that the response slabs are centered in a cubic unit-cell of dimension $d$, and each slab, $i$, with thickness, $s_i$, is associated with an electric or magnetic response whose unit-cell-averaged susceptibility, ${{\bar \chi}_i }$, is described by a Lorentzian dispersion,
\begin{equation}
\label{lorentz}
{{\bar \chi }_i}\left( f \right) =  {\chi _i^0}\frac{{f_{i}^2}}{{f_{i}^2 + j{\delta _i}f - {f^2}}}.
\end{equation}
The Lorentzian dispersion has the usual parameters of static response, $\chi _i^0$, resonance frequency, $f_i$, and loss, $\delta_i$. In this function, we do not include the high-frequency  response ($\chi _i^\infty$) which arises from resonances above the frequency range of interest. Instead, we include any response due to above-range resonances as a separate slab, with its own thickness. The susceptibility of such a slab is the constant, $\chi _i^0$, the high resonance-frequency (i.e. low frequency) limit of (\ref{lorentz}). This treatment is motivated by the fact that above-range resonances won't necessarily have the same mode shape as any of the in-range resonances. It should be noted that we have allowed a constant magnetic susceptibility to be negative in all three of the examples analyzed in this article. As pointed out by Wood and Pendry\cite{Wood2007Metamaterials}, this can be allowed in causal media. Below we highlight the one case where causality has been compromised in a specific fitted model.

The unit-cell-averaged susceptibility is defined in relation to the local, slab susceptibility by equating the polarizabilities associated with the appropriate volumes.
\begin{subequations}
\begin{eqnarray}
{\alpha _i^e} &=& \frac{p _i}{E} = {\varepsilon _0} {\chi _i^e}{s_i}{d^2} \equiv {\varepsilon _0} {\bar \chi _i^e}{d^3}\\
{\alpha_i^m} &=& \frac{m_i}{H} = {\chi _i^m}s_i{d^2} \equiv  {\bar \chi _i^m} {d^3}
\end{eqnarray}
\end{subequations}
For either electric or magnetic response, we find the local, slab susceptibility is related to the unit-cell-averaged susceptibility through the thickness ratio.
\begin{equation}
\label{local}
\chi_i =  {\bar \chi_i} \frac{d}{s_i}
\end{equation}
The reason to associate the Lorentzian parameters with the unit-cell-averaged susceptibilities instead of the local, slab susceptibilities is to decouple the thickness and Lorentzian parameters. For example, in this scheme, the limit of zero thickness (with finite polarizability) can be treated without divergent Lorentzian parameters. Also, we prefer susceptibilities to permittivities and permeabilities since their superposed response can be calculated by direct summation. The four models we consider are described below. We begin with the most general, with each Lorentzian slab allowed its own unique thickness. The other models are special cases that follow from the general one: a single thickness for all slabs, all slabs of zero thickness, and all slabs of unit cell thickness, (i.e. the homogeneous model).

\subsection{Slabs of multiple thicknesses}
The superposition of the susceptibilities of $N$ overlapping slabs centered in the unit cell, creates up to $2N+1$ distinct, contiguous, uniform material layers, including two vacuum layers if the thickest slab is thinner than the unit cell dimension. The refractive index and impedance of these layers is given by
\begin{subequations}
\label{nzMS}
\begin{eqnarray}
{n_j} &=& \sqrt {\left( {1 + \sum\limits_{i \in j} {{{\bar \chi }_i^e}\frac{d}{{{s_i}}}} } \right)\left( {1 + \sum\limits_{i \in j} {{{\bar \chi }_i^m}\frac{d}{{{s_i}}}} } \right)}\\
{z_j} &=& \sqrt {\frac{{1 + \sum\limits_{i \in j} {{{\bar \chi }_i^m}\frac{d}{{{s_i}}}} }}{{1 + \sum\limits_{i \in j} {{{\bar \chi }_i^e}\frac{d}{{{s_i}}}} }}}
\end{eqnarray}
\end{subequations}
where the index, $i$, refers to the slabs, and the index, $j$, refers to the layers. The notation, $i \in j$, means sum over all of the slabs, $i$, that overlap in the layer, $j$.  We assume these contiguous layers are ordered from front-to-back. The \textit{front} of the unit cell is the reference plane for the reflection coefficient, $r$,and the transmission coefficient, $t'$, refers to complex plane-wave amplitude at the \textit{back} of the unit relative to the \textit{front}, following the notation of Smith. We can find these coefficients using transfer matrix methods. In particular the transfer matrix of the individual layers is given by
\begin{subequations}
\label{tjmatrix}
\begin{eqnarray}
{{\mathbf{T}}_j} &=& \left( {\begin{array}{*{20}{c}}
{{\alpha _j} - {i\beta _j}\zeta _j^ + }&{ - {i\beta _j}\zeta _j^ - }\\
{{i\beta _j}\zeta _j^ - }&{{\alpha _j} + {i\beta _j}\zeta _j^ + }
\end{array}} \right)\\
{\alpha _j} &=& \cos \left( {{n_j}k{l_j}} \right)\\
{\beta _j} &=& \sin \left( {{n_j}k{l_j}} \right)\\
\zeta _j^ \pm  &=& \frac{1}{2}\left( {\frac{1}{{{z_j}}} \pm {z_j}} \right)
\end{eqnarray}
\end{subequations}
where $l_j$ is the thickness of a layer and $k$ is the free-space wave-vector. The transfer matrix of the combined set of layers just the matrix product of the individual- layer transfer matrices,
\begin{equation}
\mathbf{T} = \prod \limits_{\it j} \mathbf{T}_{\it j}
\end{equation}
and the reflection and transmission coefficients are found simply from the elements of the combined transfer matrix.
\begin{subequations}
\label{coeffsMS}
\begin{eqnarray}
r &=& \frac{T_{21}}{T_{11}}
\\
t' &=& \frac{1}{T_{11}} 
\end{eqnarray}
\end{subequations}
Unless there is just one slab, it is clear that equations (\ref{coeffsMS}) cannot be inverted to find the material properties of the slabs as a function of the reflection and transmission coefficients independently at each frequency point, since the number of unknowns exceeds the number of equations. Instead, to find the slab material properties from simulation derived reflection and transmission coefficients, we perform a least-squares fit over a \textit{range} of frequencies, employing a distinct Lorentzian dispersion for each slab\cite{Lubkowski2007Extraction}. There are two or four fit parameters per slab depending on whether the slab represents an above-range resonance or an in-range resonance, respectively. The full set of fit parameters is
\begin{equation}
\begin{matrix}
  {s_1},\space\chi _1^0{\rm{\hspace{10 mm}or \hspace{10 mm}   }}{s_1},\space\chi _1^0,\space{f_1},\space{\delta _1}\\
{s_2},\space\chi _2^0{\rm{\hspace{10 mm}or \hspace{10 mm}   }}{s_2},\space\chi _2^0,\space{f_2},\space{\delta _2} \\
  \vdots \\
{s_N},\space\chi _N^0{\rm{\hspace{10 mm}or \hspace{10 mm}   }}{s_N},\space\chi _N^0,\space{f_1},\space{\delta _N}
 \end{matrix}
\end{equation}
If the number of slabs is appropriate to the number of resonances in the range of frequencies used, there is sufficient information in the reflection and transmission data to precisely determine all of these parameters.

\subsection{Slabs of a single finite thickness}
\begin{figure}
\includegraphics[width = 86 mm]{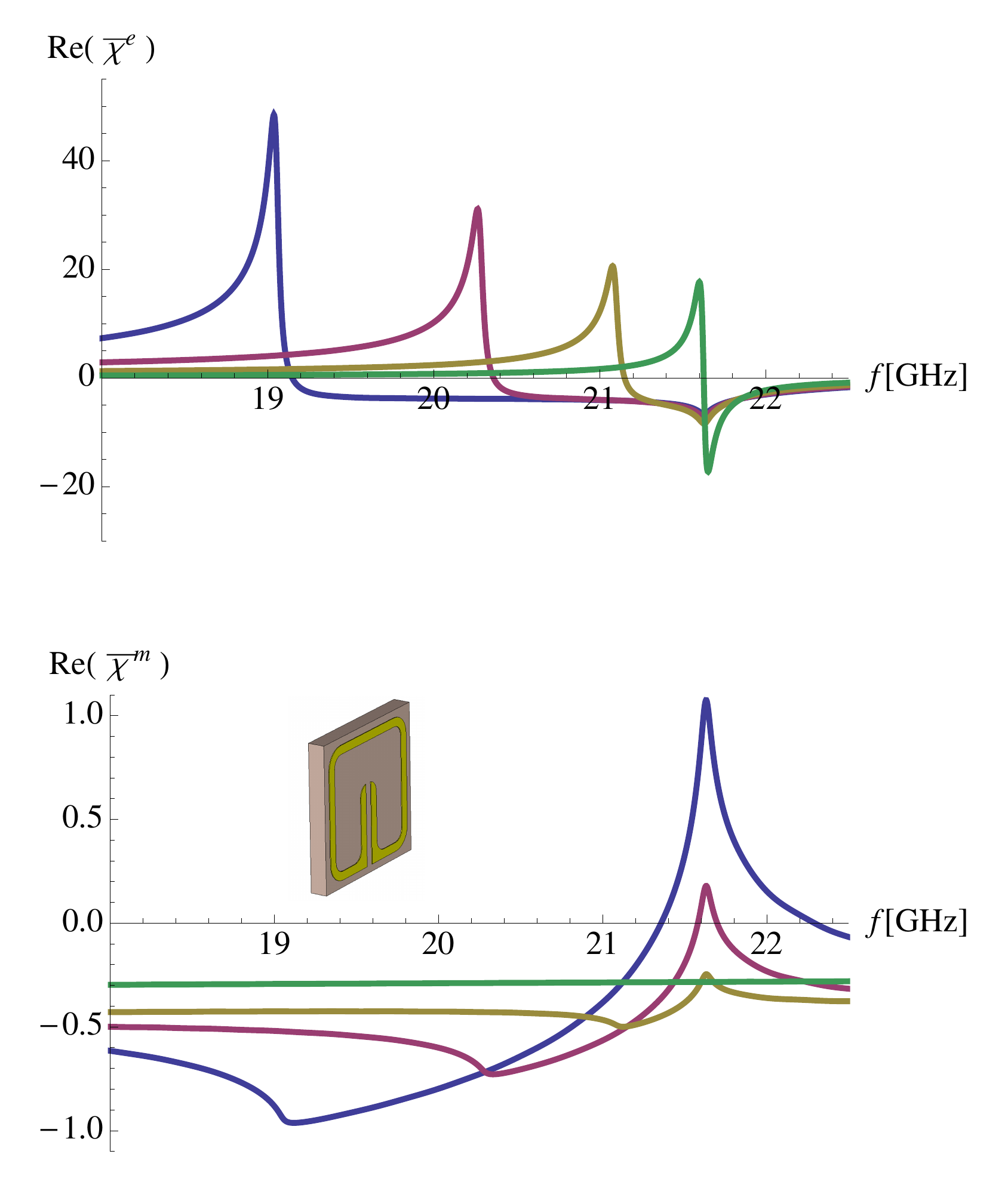}
\caption{\label{tuneFig}Tuning the model slab thickness, $s$, to a physically preferred value that minimizes non-causal response. The unit cell is a split ring resonator (microwave cloak, design 1, inset). The frequency range shown includes the lowest electric resonance (above the lowest magnetic resonance). The blue, red, yellow and green curves are for $s$ = 3.33 (full unit cell), 2.54, 1.74 and 0.95 mm, respectively. At the preferred thickness (green curve) the permittivity approaches the Lorentzian form and the “anti-resonance” response of the permeability has been eliminated. Note that for thicknesses thinner than the preferred value, non-physical features return to the response (not shown). }
\end{figure}

The single slab is the same as the multi-slab, except all of the Lorentzian response functions are associated with a single thickness, $s$. The refractive index and impedance for the single material layer that is coincident with this slab is a simplification of equations (\ref{nzMS}),
\begin{subequations}
\label{nzSS}
\begin{eqnarray}
n &=& \sqrt {\left( {1 + \frac{d}{s}\sum\limits_i {{{{\bar \chi }_i^e}}} } \right)\left( {1 + \frac{d}{s}\sum\limits_i {{{{\bar \chi }_i^m}}} } \right)},\\
z &=& \sqrt {\frac{{1 + \frac{d}{s}\sum\limits_i {{{{\bar \chi }_i^m}}} }}{{1 + \frac{d}{s}\sum\limits_i {{{{\bar \chi }_i^e}}} }}}.
\end{eqnarray}
\end{subequations}
In addition to this slab, the unit-cell is comprised of two surrounding, free-space layers of thickness $(d-s)/2$. It is convenient to define reflection and transmission coefficients, $r_s$ and $t'_s$, that are referenced (de-embedded) to the front and back of the slab.
\begin{subequations}
\label{deemb}
\begin{eqnarray}
r &=& {e^{ - jk\left( {d - s} \right)}}{r_s}\\
t' &=& {e^{ - jk\left( {d - s} \right)}}{t'_s}
\end{eqnarray}
\end{subequations}
These new coefficients can be found from a single transfer matrix, which results in the frequently used expressions\cite{Smith2002Determination}.
\begin{subequations}
\label{rstps}
\begin{eqnarray}
\frac{1}{{{{t'}_s}}} &=& \cos \left( {nks} \right) + \frac{1}{2}j\left( {z + \frac{1}{z}} \right)\sin \left( {nks} \right)\\
\frac{{{r_s}}}{{{{t'}_s}}} &=& \frac{1}{2}j\left( {z - \frac{1}{z}} \right)\sin \left( {nks} \right)
\end{eqnarray}
\end{subequations}
The equations (\ref{rstps}) can, of course, be inverted to yield
\begin{subequations}
\label{nzSS2}
\begin{eqnarray}
n &=& \frac{1}{{ks}}{\cos ^{ - 1}}\left[ {\frac{{1 - r_s^2 + {t'}_s^2}}{{2{{t'}_s}}}} \right]\\
z &=& \sqrt {\frac{{{{\left( {1 + {r_s}} \right)}^2} - {t'}_s^2}}{{{{\left( {1 - {r_s}} \right)}^2} - {t'}_s^2}}}
\end{eqnarray}
\end{subequations}
so that the slab material properties can be found from simulation derived reflection and transmission coefficients, independently at each frequency.  We can also find the unit-cell-averaged susceptibilities.
\begin{subequations}
\label{chibarSS}
\begin{eqnarray}
{{\bar \chi }^e} &=& \left(\ \frac{n}{z} - 1 \right)\frac{s}{d}\\
{{\bar \chi }^m} &=& \left(\ nz - 1 \right)\frac{s}{d}
\end{eqnarray}
\end{subequations}
The slab material properties, (\ref{nzSS}), as well as unit-cell-averaged susceptibilities, (\ref{chibarSS}), are dependent on the value chosen for the thickness, $s$. One can choose this thickness to minimize spatial-dispersion induced artifacts. An example is shown in Fig.\ref{tuneFig}.

\subsection{Thin sheet: Slabs of infinitesimal thickness}
Consider the case of a super-position of slabs, all of infinitesimal thickness, but where we assume the unit-cell averaged susceptibilities are finite\cite{smith2010Analytic,Fietz2010Homogenization}.  To take the thin-slab limit of equations (\ref{deemb}), (\ref{rstps}) and (\ref{nzSS2}), we require the following limits, which we find using equations (\ref{nzSS}).
\begin{subequations}
\label{zero1}
\begin{eqnarray}
\mathop {\lim }\limits_{s \to 0} ns &=& \sqrt {\left(\sum\limits_i {{{\bar \chi }_{ei}}}\right)\left( \sum\limits_i {{{\bar \chi }_{mi}}}\right) }\; d \equiv {n_0}d\\
\mathop {\lim }\limits_{s \to 0} z &=& \sqrt {\frac{{\sum\limits_i {{{\bar \chi }_{mi}}} }}{{\sum\limits_i {{{\bar \chi }_{ei}}} }}}  \equiv {z_0}
\end{eqnarray}
\end{subequations}
As seen from (\ref{local}), the local susceptibilities are divergent, so that the unit constant in (\ref{nzSS}) can been neglected by comparison. The limits are, however, finite. On the right-hand-side we define special index, $n_0$, and impedance, $z_0$, for this zero-thickness layer.
We find the relationship between the thin slab coefficients $r_0$ and $t'_0$ and the de-embedded coefficients $r$ and $t'$ by taking the same limit of equation (\ref{deemb})
\begin{subequations}
\begin{eqnarray}
r &=& {e^{ - jk{d}}}{r_0}\\
t' &=& {e^{ - jk{d}}}{t'_0}
\end{eqnarray}
\end{subequations}
Additionally, taking the limit of equations (\ref{rstps}) and (\ref{nzSS2}) we find
\begin{subequations}
\begin{eqnarray}
\frac{1}{{{{t'}_0}}} &=& \cos \left( {{n_0}kd} \right) + \frac{1}{2}j\left( {{z_0} + \frac{1}{{{z_0}}}} \right)\sin \left( {{n_0}kd} \right)\\
\frac{{{r_0}}}{{{{t'}_0}}} &=& \frac{1}{2}j\left( {{z_0} - \frac{1}{{{z_0}}}} \right)\sin \left( {{n_0}kd} \right)
\end{eqnarray}
\end{subequations}
and
\begin{subequations}
\label{zero4}
\begin{eqnarray}
{n_0} &=& \frac{1}{{kd}}{\cos ^{ - 1}}\left[ {\frac{{1 - r_0^2 + {t'}_0^2}}{{2{{t'}_0}}}} \right]\\
{z_0} &=& \sqrt {\frac{{{{\left( {1 + {r_0}} \right)}^2} - {t'}_0^2}}{{{{\left( {1 - {r_0}} \right)}^2} - {t'}_0^2}}}
\end{eqnarray}
\end{subequations}
We can also include infinitesimal slabs in a multi-thickness slab. All of the overlapping infinitesimal slabs will contribute a single transfer matrix, $\mathbf{T}_j$. This transfer matrix is found by taking the $l_j \rightarrow 0$ limit of equations (\ref{tjmatrix}), just as we took the limit of equations (\ref{rstps}) and (\ref{nzSS2}). Any finite slabs that overlap the infinitesimal slabs do not contribute and can be neglected for this transfer matrix.

\subsection{Homogenous: Slabs all of thickness $d$}
The standard model for material property extraction is the uniform or homogeneous medium. The primary advantage of this model is simplicity. Engineers already know how to design and analyze devices that incorporate uniform or slowly varying media. Like the single-thickness and thin sheet models, the homogeneous model can be inverted, and the inverse formulas applied independently at each frequency point. Equations (\ref{nzSS}-\ref{nzSS2}) can be applied by setting $s=d$.

Obviously, these material properties will precisely represent, with the use of the Fresnel formulas (\ref{rstps}), the reflection and transmission behavior for a sample of the same thickness from which they were extracted (though not very usefully). They may also precisely predict the reflection and transmission behavior for an arbitrary sample thickness, if the inter-cell, electromagnetic coupling is restricted to dipole interactions. Also, if a sufficient number of unit cells are used in the  propagation direction, the material properties will converge to those of  a bulk medium, regardless of the coupling behavior. The extracted bulk properties will apply to all sufficiently thick slabs.

Unfortunately, these material properties will usually provide limited physical insight, because the model is often too simple to represent the metamaterial in question. Wedging the reflection and transmission behavior into this over-simplified model leads to material properties that violate physical principles. Specifically, the practical, moderate values of free-space phase advance across the unit cell (i.e. $\lambda_0/d$), together with the extreme polarization found near a resonance, lead to significant field variation across the unit cell, which is not compatible with an effective, uniform (spatial dispersion free) medium. The resulting material properties are then a poor fit to the simple Lorentzian line shapes one expects from an isolated, second-order resonance\cite{Koschny2005Impact,Alitalo2013Experimental}.

\section{Examples}

\begin{figure}
\includegraphics[width = 86 mm]{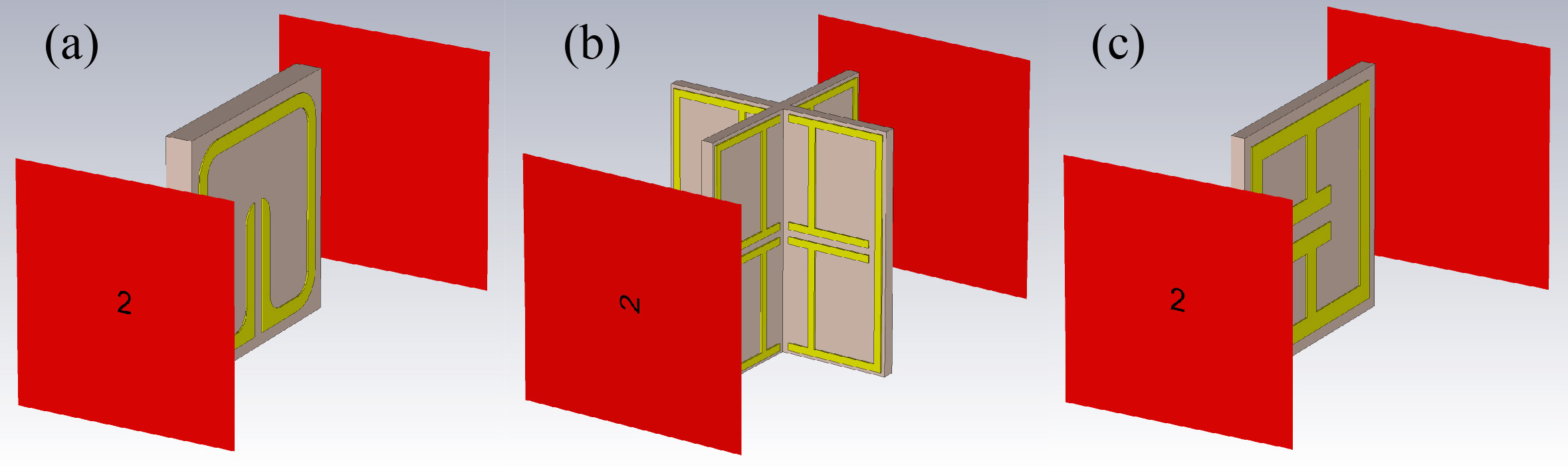}
\caption{\label{unitcells}The three unit cells analyzed: (a) the microwave invisibility cloak, cylinder 1 design, (b) a two-dimensional negative index medium design, (c) the original ELC resonator design. }
\end{figure}

We demonstrate the fitting of the four models to simulation derived reflection and transmission coefficients (S-parameters) for two well known unit cells: the microwave invisibility cloak (Fig.\ref{unitcells}a)\cite{Schurig2006Metamaterial} and the ELC resonator (Fig.\ref{unitcells}c)\cite{Schurig2006Electric}. Another unit cell, one designed for two-dimensional, isotropic negative index is also included (Fig.\ref{unitcells}b). This latter design provides a somewhat more complex response with three overlapping resonances. 

\subsection{Microwave Cloak Unit Cell}
\begin{figure*}
\includegraphics[width = 165 mm]{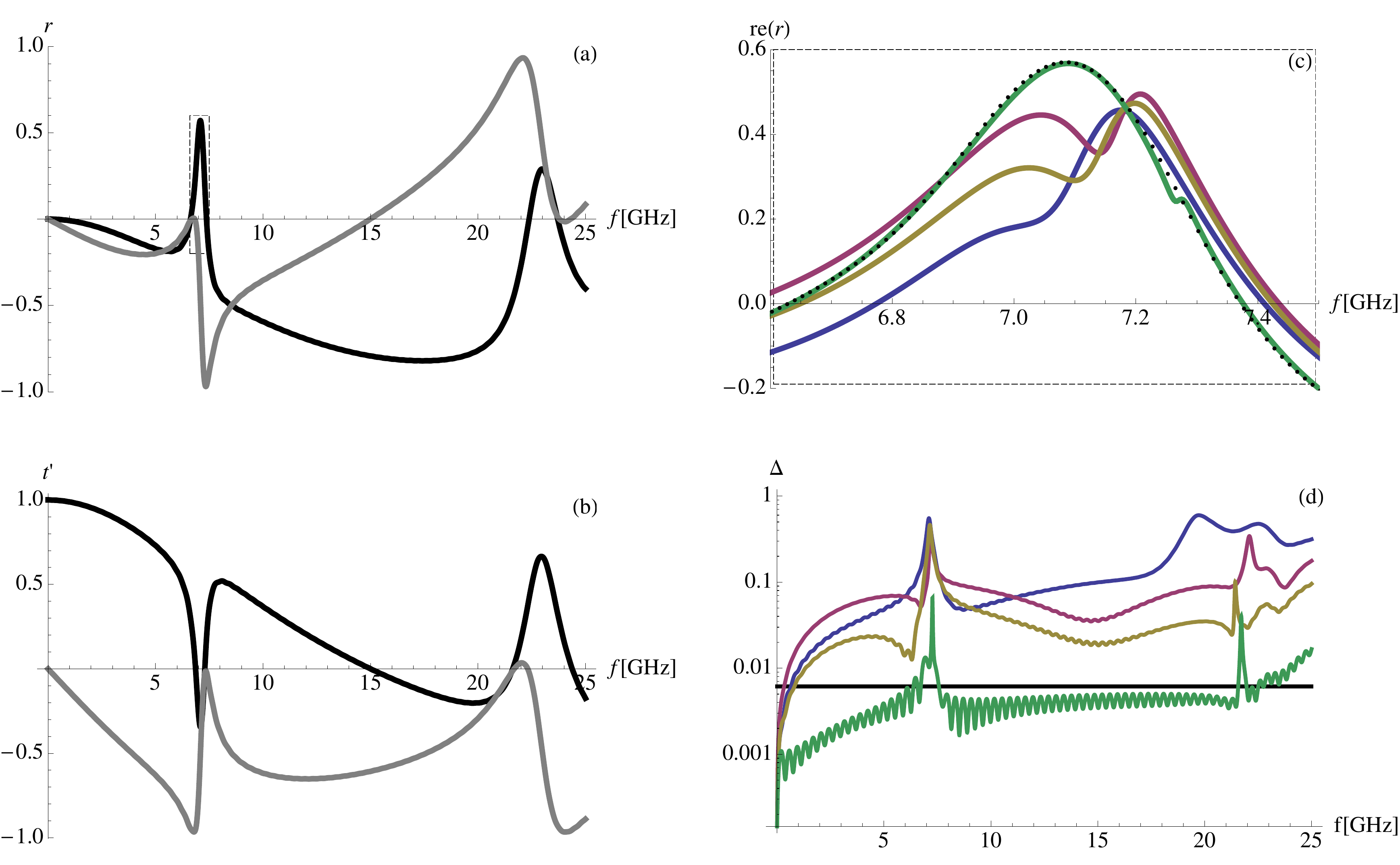}
\caption{\label{SRR}Microwave cloak unit cell. From simulation, the real and imaginary part of the (a) reflection coefficient and (b) transmission coefficient. (c) Best global fits of the four models to the simulated data (black dots). The narrow range of frequencies displayed is indicated by the dashed box in (a). The models are shown in: green (multi-thickness), yellow (single thickness), red (thin sheet) and blue (homogeneous). (d) The combined residuals, $\Delta$, as defined in equation (\ref{delta}), with the same color scheme.}
\end{figure*}

Fitting of all the models provides good constraint of the fit parameters, as seen from the parameter uncertainties in Table \ref{tab:SRR}. However, only the multi-thickness model provides a quantitatively precise fit. In plots covering the entire fitted frequency range (such as Fig. \ref{SRR}(a) and (b)), simulation and model curves would be indistinguishable. The small deviations between model and data can only be observed in zoomed-in sections of the plot (Fig. \ref{SRR}(c)) or in plots of the residuals (Fig. \ref{SRR}(d)). The chi-squared per degree-of-freedom and probability measures of model appropriateness give quantitative support to choosing the multi-thickness model. (These measures are normalized to the multi-thickness model fit, as described in section IV.) The probabilities for the other models are so small that they are essentially zero.

One subtlety arose in obtaining the best fits for the multi-thickness model. As can be seen in Table \ref{tab:SRR}, three of the four thicknesses are given as approximately zero. Here the results for a zero thickness slab (equations (\ref{zero1})-(\ref{zero4})) were not used. Instead, a \textit{different} small but \textit{finite} thickness was used for each slab, with $s_1<s_3<s_4$. The quality of the fit was independent of these thickness for any so-ordered set of values less than about $0.01$mm. Good, but less impressive, fits could be obtained using a zero-thickness slab in the multi-thickness model. 

\begin{table*}
\caption{\label{tab:SRR}Microwave cloak unit cell.}
\begin{ruledtabular}
\begin{tabular}{lvvuuttss}
& \multicolumn{2}{z}{multi-thickness} & \multicolumn{2}{y}{single thickness} & \multicolumn{2}{x}{thin sheet} & \multicolumn{2}{w}{homogeneous} \\
$\chi^2/\mathop{\rm DOF}$& \multicolumn{2}{v}{1} & \multicolumn{2}{u}{80} & \multicolumn{2}{t}{200}  & \multicolumn{2}{s}{1000} \\
probability& \multicolumn{2}{v}{1} & \multicolumn{2}{u}{\sim0} & \multicolumn{2}{t}{\sim0}  & \multicolumn{2}{s}{\sim0} \\
parameters & \multicolumn{2}{v}{9} &  \multicolumn{2}{u}{9} &  \multicolumn{2}{t}{8} &  \multicolumn{2}{s}{8} \\
\hline
\multicolumn{9}{l}{electric parameters}\\
\hline
$s_1$ & \sim0 & & 1.2091 & \pm 0.0004 & 0 & & d & \\
$\chi_1^0$ & 0.31754 & \pm 0.00004 & 0.36051 & \pm 0.00006 & 0.26109 & \pm 0.00007 & 0.61682 & \pm 0.00007 \\
$f_1$ & 21.6825 & \pm 0.0002 & 21.4801 & \pm 0.0002 & 21.9129 & \pm 0.0003 & 20.3653 & \pm 0.0003 \\
$\delta_1$ & 0.0506 & \pm 0.0002 & 0.0587 & \pm 0.0002 & 0.1911 & \pm 0.0002 & 0.821 & \pm 0.0002 \\
\hline
$s_2$ & 2.6872 & \pm 0.0008 & 1.2091 & \pm 0.0004 & 0 & & d & \\
$\chi_2^0$ & 1.6023 & \pm 0.0002 & 1.4031 & \pm 0.0001 & 1.2887 & \pm 0.0001 & 1.429 & \pm 0.0001 \\
\hline
\multicolumn{9}{l}{magnetic parameters}\\
\hline
$s_3$ & \sim0 & & 1.2091 & \pm 0.0004 & 0 & & d & \\
$\chi_3^0$ & 0.29928 & \pm 0.00005 & 0.2937 & \pm 0.00005 & 0.31041 & \pm 0.00005 & 0.26302 & \pm 0.00005 \\
$f_3$ & 7.27931 & \pm 0.00005 & 7.18931 & \pm 0.00005 & 7.20405 & \pm 0.00005 & 7.15342 & \pm 0.00005 \\
$\delta_3$ & 0.02444 & \pm 0.00009 & 0.09179 & \pm 0.00008 & 0.0712 & \pm 0.00009 & 0.11427 & \pm 0.00009 \\
\hline
$s_4$ & \sim0 & & 1.2091 & \pm 0.0004 & 0 & & d & \\
$\chi_4^0$ & -0.30974 & \pm 0.00006 & -0.26398 & \pm 0.00004 & -0.25707 & \pm 0.00004 & -0.42512 & \pm 0.00004 \\
\end{tabular}
\end{ruledtabular}
\end{table*}

\subsection{2D Isotropic Unit Cell}
As with the cloak unit cell, here only the multi-thickness model provides a quantitatively precise fit as seen from the chi-squared per degree-of-freedom and probability measures of model appropriateness in Table \ref{tab:2Diso}. Particular to this unit cell, the homogeneous model obtains the same quality of fit while omitting the second electric resonance included in the other models. This is more an indication of a poor fit, than evidence for the absence of the resonance. Though the multi-thickness model includes fourteen parameters---a substantial number---the complexity of the four, real-valued data sets (i.e. the real and imaginary parts of the reflection and transmission curves) is such that only a physically-motivated, and correct model has any chance of providing a good fit. This is born out in the probability measure, which includes an Occam’s penalty factor for each parameter (as describe in section IV). Here the zero-valued thicknesses in the multi-thickness model \textit{do} use the limiting results of the zero thickness slab, equations (\ref{zero1})-(\ref{zero4}).

\begin{figure*}
\includegraphics[width = 178 mm]{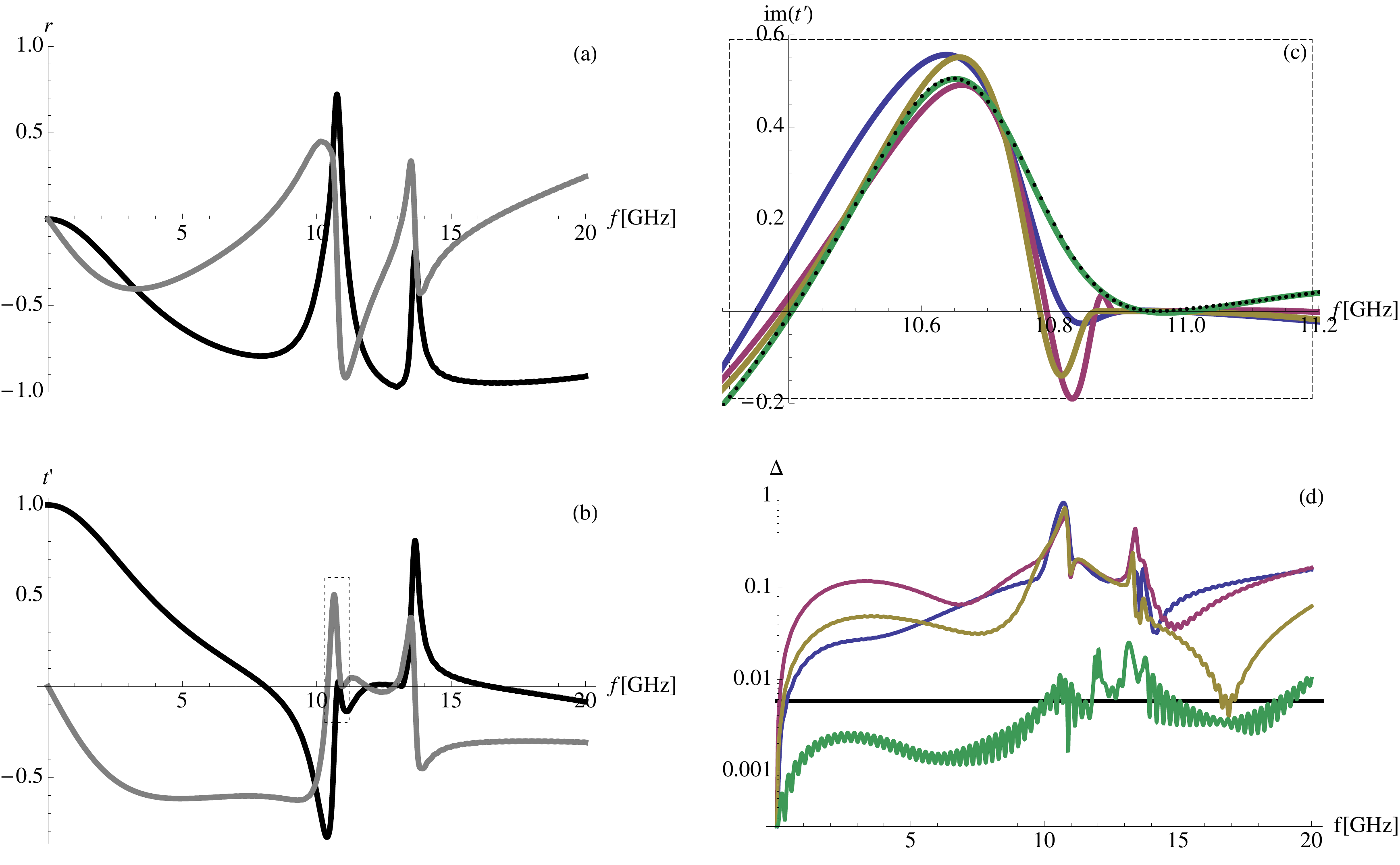}
\caption{\label{2Diso}2D isotropic negative index unit cell. From simulation, the real and imaginary part of the (a) reflection coefficient and (b) transmission coefficient. (c) Best global fits of the four models to the simulated data (black dots). The narrow range of frequencies displayed is indicated by the dashed box in (a). The models are shown in: green (multi-thickness), yellow (single thickness), red (thin sheet) and blue (homogeneous). (d) The combined residuals, $\Delta$, as defined in equation (\ref{delta}), with the same color scheme.}
\end{figure*}

\begin{table*}
\caption{\label{tab:2Diso}2D isotropic negative index unit cell.}
\begin{ruledtabular}
\begin{tabular}{lvvuuttss}
& \multicolumn{2}{z}{multi-thickness} & \multicolumn{2}{y}{single thickness} & \multicolumn{2}{x}{thin sheet} & \multicolumn{2}{w}{homogeneous} \\
$\chi^2/\mathop{\rm DOF}$ & \multicolumn{2}{v}{1} & \multicolumn{2}{u}{500} & \multicolumn{2}{t}{700}  & \multicolumn{2}{s}{700} \\
probability & \multicolumn{2}{v}{1} & \multicolumn{2}{u}{\sim0} & \multicolumn{2}{t}{\sim0}  & \multicolumn{2}{s}{\sim0} \\
parameters & \multicolumn{2}{v}{14} &  \multicolumn{2}{u}{11} &  \multicolumn{2}{t}{11} &  \multicolumn{2}{s}{8} \\
\hline
\multicolumn{9}{l}{electric parameters}\\
\hline
$s_1$ & 0 & & 1.4122 & \pm 0.0004 & 0 & & d & \\
$\chi_1^0$ & 0.321 & \pm 0.0006 & 0.4367 & \pm 0.0001 & 0.2938 & \pm 0.00008 & 0.5606 & \pm 0.0001 \\
$f_1$ & 12.8644 & \pm 0.0004 & 13.2675 & \pm 0.00009 & 13.3235 & \pm 0.00008 & 13.2372 & \pm 0.00009 \\
$\delta_1$ & 0.065 & \pm 0.0001 & 0.06952 & \pm 0.00007 & 0.10551 & \pm 0.00009 & 0.05856 & \pm 0.00007 \\
\hline
$s_2$ & 0 & & 1.4122 & \pm 0.0004 & 0 & & - & \\
$\chi_2^0$ & 0.7124 & \pm 0.0007 & 0.342 & \pm 0.0003 & 0.748 & \pm 0.0004 & - & \\
$f_2$ & 10.8788 & \pm 0.0001 & 9.896 & \pm 0.0006 & 9.9264 & \pm 0.0007 & - & \\
$\delta_2$ & 0.0547 & \pm 0.0001 & 1.315 & \pm 0.001 & 3.344 & \pm 0.003 & - & \\
\hline
$s_3$ & 2.2578 & \pm 0.0005 & 1.4122 & \pm 0.0004 & 0 & & d & \\
$\chi_3^0$ & 7.4204 & \pm 0.0007 & 6.9715 & \pm 0.0006 & 5.7128 & \pm 0.0004 & 8.2337 & \pm 0.0005 \\
\hline
\multicolumn{9}{l}{magnetic parameters}\\
\hline
$s_4$ & 0.264 & \pm 0.002 & 1.4122 & \pm 0.0004 & 0 & & d & \\
$\chi_4^0$ & 0.3748 & \pm 0.0001 & 0.32425 & \pm 0.00004 & 0.40891 & \pm 0.00004 & 0.26485 & \pm 0.00003 \\
$f_4$ & 11.9281 & \pm 0.0009 & 10.8916 & \pm 0.00005 & 10.9127 & \pm 0.00005 & 10.9148 & \pm 0.00005 \\
$\delta_4$ & 0.0691 & \pm 0.0001 & 0.06142 & \pm 0.00006 & 0.02399 & \pm 0.00006 & 0.14302 & \pm 0.00007 \\
\hline
$s_5$ & 0.044 & \pm 0.004 & 1.4122 & \pm 0.0004 & 0 & & d & \\
$\chi_5^0$ & -0.4018 & \pm 0.0003 & -0.3861 & \pm 0.00007 & -0.5123 & \pm 0.0001 & -0.47033 & \pm 0.00005 \\
\end{tabular}
\end{ruledtabular}
\end{table*}

\subsection{ELC Unit Cell}
For this unit cell, Fig.\ref{unitcells}(c), the results are much the same as the other two, in terms of the quality of the model fits. However, we point out two problems that occurred with the multi-thickness model fit, as shown in Table \ref{tab:ELC}. One is the that thickness of the constant magnetic susceptibility slab, $s_3$, exceeds the dimension of the unit cell, $d$, and all of its material components. This seems to be a non-physical result, but perhaps a generous interpretation of \textit{effective} polarization would allow this. The second problem is more serious. Away from the unit-cell center, where only the constant magnetic susceptibility slab is present, the wave speed exceeds the speed of light in vacuum, $c$. This is the only case where this violation is found. In all other models, and all other unit-cells analyzed, the frequency independent response supports only evanescent waves or waves with speeds less than $c$, or has zero thickness. When this is the case, the frequency independent response is causal, and it remains so when combining with a causal Lorentzian response. It may be that the fit parameters could be constrained to prevent causality violations, and compelling fits still obtained, though that was not confirmed here.
\begin{figure*}
\includegraphics[width = 178 mm]{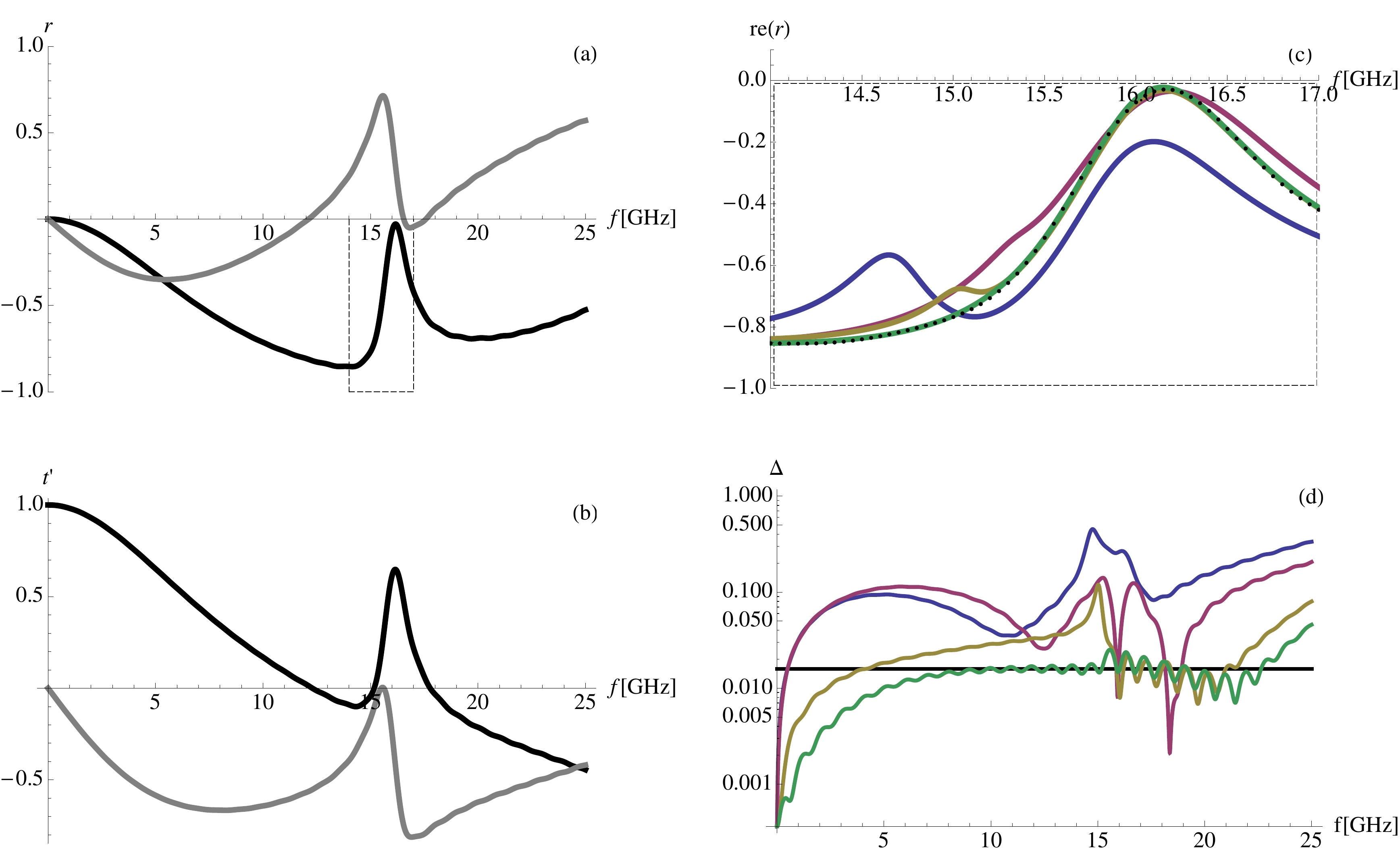}
\caption{\label{ELC}ELC unit cell. From simulation, the real and imaginary part of the (a) reflection coefficient and (b) transmission coefficient. (c) Best global fits of the four models to the simulated data (black dots). The narrow range of frequencies displayed is indicated by the dashed box in (a). The models are shown in: green (multi-thickness), yellow (single thickness), red (thin sheet) and blue (homogeneous). (d) The combined residuals, $\Delta$, as defined in equation (\ref{delta}), with the same color scheme.}
\end{figure*}

\begin{table*}
\caption{\label{tab:ELC}ELC unit cell.}
\begin{ruledtabular}
\begin{tabular}{lvvuuttss}
& \multicolumn{2}{z}{multi-thickness} & \multicolumn{2}{y}{single thickness} & \multicolumn{2}{x}{thin sheet} & \multicolumn{2}{w}{homogeneous} \\
$\chi^2/\mathop{\rm DOF}$& \multicolumn{2}{v}{1} & \multicolumn{2}{u}{4} & \multicolumn{2}{t}{40}  & \multicolumn{2}{s}{100} \\
probability & \multicolumn{2}{v}{1} & \multicolumn{2}{u}{\sim0} & \multicolumn{2}{t}{\sim0}  & \multicolumn{2}{s}{\sim0} \\
parameters & \multicolumn{2}{v}{6} &  \multicolumn{2}{u}{6} &  \multicolumn{2}{t}{5} &  \multicolumn{2}{s}{5} \\
\hline
\multicolumn{9}{l}{electric parameters}\\
\hline
$s_1$ & 0 & & 1.7125 & \pm 0.0009 & 0 & & d & \\
$\chi_1^0$ & 0.3909 & \pm 0.0001 & 0.4316 & \pm 0.0001 & 0.33828 & \pm 0.0001 & 0.5388 & \pm 0.0002 \\
$f_1$ & 15.3043 & \pm 0.0003 & 15.156 & \pm 0.0003 & 15.2831 & \pm 0.0003 & 15.0181 & \pm 0.0003 \\
$\delta_1$ & 0.2651 & \pm 0.0004 & 0.2767 & \pm 0.0003 & 0.3432 & \pm 0.0004 & 0.4076 & \pm 0.0003 \\
\hline
$s_2$ & 0 & & 1.7125 & \pm 0.0009 & 0 & & d & \\
$\chi_2^0$ & 1.872 & \pm 0.0005 & 2.279 & \pm 0.0004 & 1.798 & \pm 0.0002 & 2.8887 & \pm 0.0004 \\
\hline
\multicolumn{9}{l}{magnetic parameters}\\
\hline
$s_3$ & 5.114 & \pm 0.003 & 1.7125 & \pm 0.0009 & 0 & & d & \\
$\chi_3^0$ & -0.8308 & \pm 0.0009 & -0.25693 & \pm 0.00008 & -0.26243 & \pm 0.0001 & -0.43796 & \pm 0.00006 \\
\end{tabular}
\end{ruledtabular}
\end{table*}

\section{Statistical Model Selection}
To compare different models for a given data set, one must quantify the goodness of fit as well as assess a penalty for each free fit parameter. This penalty is sometimes called the \textit{Occam's factor}. An expression for the posterior probability of model correctness can be found using Bayes’ Theorem. We follow the notation and conventions of the book by Sivia and Skilling\cite{Sivia2006Data}.
\begin{equation}
{\mathop{\rm prob}\nolimits} \left( {M\left| D \right.} \right) = \frac{{{\mathop{\rm prob}\nolimits} \left( {D\left| M \right.} \right){\mathop{\rm prob}\nolimits} \left( M \right)}}{{{\mathop{\rm prob}\nolimits} \left( D \right)}}
\end{equation}
This expression gives the posterior probability that the model, $M$, is correct given the data, $D$. The required factors include: the probability of the data given the model (also called the likelihood function), the prior probability for the model, and the probability of the data. The last is usually not directly calculable, since it would require integrating over all possible models. The need for it can be eliminated by using a normalization condition, or by seeking only a ratio for model-correctness probabilities between two alternative models. Here the data, $D$, refers to the reflection and transmission coefficients from a simulation. If we assume that the log of the parameter dependent likelihood function can be well approximated by a quadratic series expansion around the best fit parameter values, $\bm{\lambda}_0$, and use a uniform probability over a finite interval for the prior parameter values, we find the parameter dependent likelihood function to be,
\begin{equation}
\label{likeli}
{\mathop{\rm prob}\nolimits} \left( {D|M} \right) = {\mathop{\rm prob}\nolimits} \left( {D|\bm{\lambda}_0,M} \right)\prod\limits_i {\frac{{\sqrt {2\pi } \delta {\lambda _i}}}{{\Delta {\lambda _i}}}}
\end{equation}
where the $\delta\lambda_i$ are the parameter uncertainties and the $\Delta\lambda_i$ are the prior parameter ranges. The product of factors on the right comprise the \textit{Occam's penalty} for adding free parameters. (The series approximation is the same used in  the standard calculation of the covariance matrix.) The ratio of model correctness probabilities for two models, $M_1$ and $M_2$ is then
\begin{equation}
\frac{{{\mathop{\rm prob}\nolimits} \left( {{M_1}\left| D \right.} \right)}}{{{\mathop{\rm prob}\nolimits} \left( {{M_2}\left| D \right.} \right)}} = \frac{{{\mathop{\rm prob}\nolimits} \left( D \mid \bm\lambda_0,M_1 \right)\prod\limits_i {\frac{{\sqrt {2\pi } \delta {\lambda _i}}}{{\Delta {\lambda _i}}}} }}{{{\mathop{\rm prob}\nolimits} \left( D \mid \bm\mu_0,M_2 \right)\prod\limits_j {\frac{{\sqrt {2\pi } \delta {\mu _j}}}{{\Delta {\mu _j}}}} }}
\end{equation}
where the $\lambda_i$ are the parameters for model $M_1$ and the $\mu_j$ are the parameters for model $M_2$. This assumes that the two models are equally likely prior to examining the data
\begin{equation}
\frac{{{\mathop{\rm prob}\nolimits} \left( {{M_1}} \right)}}{{{\mathop{\rm prob}\nolimits} \left( {{M_2}} \right)}} = 1
\end{equation}
The parameter uncertainties are given by the square-root of the diagonal elements of the covariance matrix.
\begin{equation}
\delta {\lambda _i} = \sqrt {{\Sigma _{ii}}}
\end{equation}
where as usual, the covariance matrix is given by the inverse of the Hessian (matrix of second derivatives) of the log of the likelihood function evaluated at the best fit parameters.
\begin{equation}
\bm \Sigma  =  - {\left( {{\nabla _{\bm \lambda }}{\nabla _{\bm \lambda }}L\left( {{{\bm \lambda }_0}} \right)} \right)^{ - 1}}
\end{equation}
For a least-squares fit, the log of the likelihood function is
\begin{equation}
L\left( {\bm \lambda } \right) = {\rm{constant}} - \frac{1}{2}{\chi ^2}\left( {\bm \lambda } \right)
\end{equation}
where $\chi^2$ is the sum of the squares of the normalized residuals
\begin{equation}
{\chi ^2}\left( {\bm \lambda } \right) = \frac{1}{{{\sigma ^2}}}\sum\limits_{k = 1}^N {{\Delta_k(\bm \lambda)}^2 }
\end{equation}
where
\begin{equation}
\label{delta}
\Delta_k(\bm \lambda) = \sqrt{{\left| r^M_k(\bm \lambda)-r_k \right|}^2+{\left| t^{\prime M}_k(\bm \lambda)-t^\prime_k \right|}^2}
\end{equation}
$k$ is a frequency index, $r^M_k(\bm \lambda)$ and $t^{\prime M}_k(\bm \lambda)$ comprise the model evaluated at frequency index $k$ and fit parameters $\bm\lambda$, and $r_k$ and $t^{\prime}_k$ are the data. We have assumed that the simulation data uncertainty, $\sigma$, is independent of frequency, and the same for all of the real and imaginary parts of $r_k$ and $t^{\prime}_k$. Note that in the parameter dependent likelihood probability
\begin{equation}
{\mathop{\rm prob}\nolimits} \left( D\mid \bm\lambda _0,M  \right) = \exp \left( {L\left( {{\bm\lambda _0}} \right)} \right) \propto \exp \left( - \frac{1}{2}\chi ^2\left( \bm\lambda _0 \right) \right)
\end{equation}
the proportionality constant does not depend on the model and may be neglected when computing the model probability ratios.

There are several issues that arise with the data uncertainty, $\sigma$, when the data is generated in a simulation. The least-squares minimization procedure is derived under the assumption that $\sigma$ is known, and describes the width of an independent Gaussian stochastic variable.
However, the error present in FDTD simulation data is all systematic and is usually dominated by finite mesh effects and oscillations introduced in the frequency domain variables by truncation of the transient variable response. We take the viewpoint that a simulation with a given mesh is a linear system with a valid response function. Thus, models may be compared using simulation data with course or fine meshes. (Of course, to best approximate a ``continuum" physical system, convergence of the response with respect to mesh density should be sought.) The remaining source of error, the transient-truncation induced oscillation, is, unfortunately, neither independent nor stochastic.

The lack of independence manifests as a correlation of the error over a significant range of frequency. For the model probability calculations, we mitigate this problem by decimating the data to a courser, evenly-spaced sub-set. The data are originally very finely-spaced in frequency, for accurate model fitting. By so decimating, We were able to reduce the (off-peak) auto-correlation of the residuals by a factor of five, while still capturing the significant behavior of the reflection and transmission coefficients. For the SRR, 2D isotropic and ELC unit-cells, the original number of frequency points was reduced from 10000, 10000, and 12500 to 121, 130, and 39 respectively. 

The lack of stochasticity means that we cannot estimate the magnitude of the error by making multiple measurements (i.e. simulations). However, due the remarkable quality of the fits using the multiple-thickness model, we can take this model to be a proxy for the unknown exact results, and use the residuals of the fits to this model to estimate $\sigma$.  With this estimate, we can test our assumption that $\sigma$ is independent of frequency, and the same for all of the real and imaginary parts of $r_k$ and $t^{\prime}_k$. We find that real and imaginary parts of $r_k$ and $t^{\prime}_k$ do indeed have very similar residuals, but the frequency independence is more questionable. Our method of estimating $\sigma$ normalizes the $\chi^2/\mathop{\rm DOF}$ and model probability to unity for the multiple-thickness model, as seen in Tables \ref{tab:SRR}, \ref{tab:2Diso} and \ref{tab:ELC}. The four models can then only be judged in a relative sense.

Finally, a slight complication arises when the off-diagonal elements of the covariance matrix are not negligible, which can result in an underestimation of the parameter uncertainties. One can correct this by finding a new set of parameters that diagonalize the covariance. Since the covariance is a real symmetric matrix, the diagonalizing parameters are given by
\begin{equation}
\bm\lambda^\prime = \mathbf{U}^T\bm\lambda
\end{equation}
where the normalized Eigenvectors of the covariance comprise the columns of the orthogonal matrix $\mathbf{U}$. The new parameter uncertainties are given by
\begin{equation}
\delta {\lambda^\prime_i} = \sqrt {\Sigma^\prime_{ii}}
\end{equation}
i.e. the square-root of the eigenvalues. We have used an upper-bound on the new a-priori parameter ranges, given by
\begin{equation}
{\Delta\lambda}_i^\prime = \sum\limits_{j} {\left|U^T_{ij}\right|{\Delta\lambda}_j}
\end{equation}
For the likelihood function, (\ref{likeli}), we had assumed that the a-priori range probabilities were uniform. This is no longer the case, but we neglect correcting for this for simplicity. The model probabilities in Tables \ref{tab:SRR}, \ref{tab:2Diso} and \ref{tab:ELC} are calculated using the diagonalized forms, but the parameter values and parameter uncertainties refer to the original, physically-derived model parameters. 

\section{Conclusion}
We have presented four models for representing the response of metamaterials to normally incident plane waves. All of these models rely on causally-responding components (with a, perhaps correctable, exception noted above). Fitted models can thus provide a comparison of unit-cell designs, using \textit{physically meaningful} figures of merit. Only one of the models---the multi-thickness model---exhibits compelling, and quantitatively precise representation of the simulated reflection and transmission behavior for the three unit-cells here analyzed. We believe this to be the simplest model thus far presented in the literature that meets the criteria of providing physically meaningful figures of merit, and quantitatively precise representation.

\section{Acknowledgments}
This work was supported by a Multidisciplinary University Research Initiative,
sponsored by the US Army Research Office (Grant No. W911NF-09-1-0539).

\bibliography{extract.bib}
\end{document}